# A Blocking Technique for Emulating Very Large Polyelectrolytes

Carsten Peterson[1], Ola Sommelius[2] and Bo Söderberg[3]

Department of Theoretical Physics, University of Lund
Sölvegatan 14A, S-223 62 Lund, Sweden

Abstract:

A new Monte Carlo method for computing thermodynamical properties of very large polyelectrolytes is presented. It is based on a renormalization group relating the original polymer to a smaller system, where in addition to the naively rescaled forces, a corrective nearest-neighbor interaction originating from the short distance Coulomb cutoff is introduced. The method is derived for low $T$ but is in the unscreened case valid for all $T$. Large polymers with $N$ monomers are emulated by Monte Carlo calculations on smaller systems, $K = N/Q$. The computational gain of the method is $Q^3$. It is explored with emphasis on room temperature. Results for $N$=10000 are presented.

---

[1] carsten@thep.lu.se
[2] ola@thep.lu.se
[3] bs@thep.lu.se

The thermodynamics of polyelectrolytes consisting of linear chains of monomers, with covalent harmonic bonding forces and Coulomb interactions, have been extensively studied with Monte Carlo Methods. Recently, high statistics results have emerged for relatively long ($N$=2048) chains [1]. A reliable measurement of e.g. the end-to-end distance $r_{ee}$ is very CPU time consuming, even when employing state-of-the-art updating methods [2, 3].

In order to circumvent the computational bottleneck, and to facilitate estimates of free energies, a variational scheme was proposed [4, 1], relying upon an Ansatz in the form of a generic Gaussian distribution. For the unscreened case the variational method overestimates $r_{ee}$ by 5-6 % at room temperature whereas for the screened case the approximation is less accurate.

In this letter we develop an alternative Monte Carlo (MC) approach, where the chains are "coarse grained" in the spirit of the renormalization group. This is accomplished by using an energy expression for blocked monomers with appropriately renormalized harmonic and Coulomb terms and an additional corrective nearest-neighbor interaction. Superficially, this is reminiscent of the well-known blob-picture of de Gennes [6], but the idea is quite different.

The approach is designed for low $T$. Indeed, at least in the absence of screening, $T_r$ turns out to be low for large $N$ [5]; thus, the approach should give good results precisely where it is needed. Our blocking scheme is derived in two ways, using (i) a real space renormalization philosophy, and (ii) a cut-off continuum formulation.

In terms of dimensionless quantities, the energy of a Coulomb chain is given by [1]

$$E = E_G + E_C = \frac{1}{2}\sum_{i=1}^{N-1} \mathbf{r}_{i,i+1}^2 + \sum_{i}\sum_{j>i} V(r_{ij}) \tag{1}$$

$$V(r) = \frac{e^{-\kappa r}}{r} \tag{2}$$

with a Boltzmann distribution $\propto \exp(-E/T)$. Here, $\mathbf{r}_{ij} = |\mathbf{x}_i - \mathbf{x}_j|$ with $\mathbf{x}_i$ as the monomer positions, while $\kappa$ is the inverse Debye screening length. With a length unit (defined by the unscreened $N$=2 equilibrium distance) of 6Å, room temperature (290K) corresponds to $T_r = 0.837808$, and 0.01M, 0.1M and 1.0M salt solution correspond to $\kappa = 0.1992, 0.630$ and $1.992$, respectively [1].

Naively, one would coarse-grain the chain into blocks of $Q$ monomers, and represent each of them by a single effective position carrying its entire charge. The remaining position variables can then be integrated out (being Gaussian), leaving a system of $K = N/Q$ effective monomers, each carrying a charge $Q$. They would be connected by harmonic bonds being a factor $Q$ weaker than the original ones.

This would be fine, were it not for the divergent short-distance behavior of the Coulomb interaction. This mainly affects neighboring blocks, where the naive effective interaction underestimates the repulsion.

This shortcoming can be remedied by a corrective interaction term between neighboring effective monomers, and we are lead to consider the following blocked energy:

$$E_B^{(K)} = \frac{1}{2Q}\sum_{i=1}^{K-1} r_{i,i+1}^2 + \frac{3}{2}(N-K)T + Q^2\sum_{i}\sum_{j>i} V(r_{ij}) + \sum_{i=1}^{K} W(r_{i,i+1}) \tag{3}$$

where $r_{ij}$ are distances between blocked monomers. The different terms represent, in order of appearance: (1) the naive effective bond energy, (2) a correction to the bond energy, to account for the bond energy of the $N$-$K$ eliminated position variables, (3) the naive effective interaction, and (4) a corrective nearest-neighbour interaction.

The correction term is determined as follows. In the low temperature limit the chain is a straight line with a slowly varying nearest-neighbour distance, which we approximate by a constant $a$. For the blocked system, the corresponding nearest-neighbour distance is then $b = Qa$. In this approximation, the true Coulomb energy is given by

$$U_N = \sum_{l=1}^{N}(N-l)V(la) \tag{4}$$

while the blocked interaction energy (including the correction term) reads

$$U_K = Q^2 \sum_{l=1}^{K}(K-l)V(lb) + KW(b) \tag{5}$$

The objective is to choose $W(b)$ such that $U_N$ and $U_K$ become identical. For the Coulomb potential of eq. (1) one obtains for $U_N$:

$$U_N = \sum_{l}^{N}(N-l)\frac{e^{-\kappa la}}{la} \approx \frac{N}{a}\sum_{l}^{\infty}\frac{e^{-\kappa la}}{l} = -\frac{N}{a}\log(1-e^{-\kappa a}) \tag{6}$$

and similarly for $U_K$:

$$U_K \approx -\frac{Q^2 K}{b}\log(1-e^{-\kappa b}) + KW(b) \tag{7}$$

where only the leading term in the sums have been included. Equating $U_N$ with $U_K$ for $b = Qa$ yields

$$W(r) = -\frac{Q^2}{r}\left[\log\left(1-e^{-\kappa r/Q}\right) - \log\left(1-e^{-\kappa r}\right)\right] \tag{8}$$

In the absence of screening ($\kappa=0$) this becomes

$$W(r) = \frac{Q^2 \log(Q)}{r} \tag{9}$$

representing an enhancement of the nearest-neighbour Coulomb interaction.

The results above are also transparent in a continuum formulation, where the ultraviolet divergence of the Coulomb interaction is regularized via an $N$-dependent cutoff, as implied by the discrete nature of the original chain. The $W$-term then corrects for the shift in the cutoff. We illustrate this in the unscreened case only.

A continuum formulation of the chain is obtained by replacing the discrete index $i$ by a continuous parameter $\sigma = i/N$, such that $\mathbf{x}_i = \mathbf{x}(\sigma)$. Then the energy of eq. (1) (in the unscreened case) transforms into:

$$E \approx \frac{1}{2N}\int_0^1 \dot{\mathbf{x}}^2(\sigma)d\sigma + N^2 \int_0^1 \int_{\sigma+1/N}^1 \frac{d\sigma d\sigma'}{|\mathbf{x}(\sigma') - \mathbf{x}(\sigma)|} \tag{10}$$

Were it not for the short-distance cutoff $(1/N)$ in $\sigma$, needed to regularize the Coulomb potential, all the $N$-dependence could be collected in a global factor $N$, by making the rescaling $\mathbf{x}(\sigma) \to N\mathbf{x}(\sigma)$. This corresponds to a mere rescaling of the temperature, and we would have the naive scaling

$$\mathbf{x}_{Qi}^{(QN,QT)} \Leftrightarrow Q\mathbf{x}_i^{(N,T)} \tag{11}$$

The $N$-dependent cutoff introduces a scale-breaking, though, and for $K < N$ the last term of eq. (10) can be rewritten as

$$N^2 \int_0^1 \int_{\sigma+1/N}^1 \frac{d\sigma d\sigma'}{|\mathbf{x}(\sigma') - \mathbf{x}(\sigma)|} = \tag{12}$$
$$N^2 \int_0^1 \int_{\sigma+1/K}^1 \frac{d\sigma d\sigma'}{|\mathbf{x}(\sigma') - \mathbf{x}(\sigma)|} + N^2 \int_0^1 \int_{\sigma+1/N}^{\sigma+1/K} \frac{d\sigma d\sigma'}{|\mathbf{x}(\sigma') - \mathbf{x}(\sigma)|}$$

In the low-temperature regime, where the chain is more or less linear, we can use the approximation

$$\mathbf{x}(\sigma') - \mathbf{x}(\sigma) \approx (\sigma' - \sigma)\dot{\mathbf{x}}(\sigma) \tag{13}$$

and obtain for the last term of eq. (12)

$$N^2 \int_0^1 \int_{\sigma+1/N}^{\sigma+1/K} \frac{d\sigma d\sigma'}{|\mathbf{x}(\sigma') - \mathbf{x}(\sigma)|} \approx N^2 \int_0^1 \int_{\sigma+1/N}^{\sigma+1/K} \frac{d\sigma d\sigma'}{(\sigma' - \sigma)|\dot{\mathbf{x}}(\sigma)|} \tag{14}$$
$$= N^2 \int_0^1 \frac{d\sigma}{|\dot{\mathbf{x}}(\sigma)|} \int_{1/N}^{1/K} \frac{d\sigma'}{\sigma'} = N^2 \log(N/K) \int_0^1 \frac{d\sigma}{|\dot{\mathbf{x}}(\sigma)|}$$

which is precisely the continuum version of the last term of eq. (3), using $W$ of eq. (9).

The correction to the naive blocked energy function was derived for low temperatures, and the method, as will be demonstrated below, indeed gives excellent results at low $T$, both for $r_{ee}$ and $E_C$, for unscreened as well as screened Coulomb interactions.

At higher temperatures, the polymer will be less linear, and it is not *a priori* certain, that the method will work there. For the *unscreened* case in the high $T$ limit, the exact results for $r_{ee}$ and $E_C$ for large $N$ can be expanded in a series in $1/T$, yielding [5]

$$<r_{ee}^2> \approx 3(N-1)T + \frac{4}{15}\sqrt{\frac{2}{\pi}} \frac{N^{5/2}}{T^{1/2}} \tag{15}$$
$$<E_C> \approx \frac{4}{3}\sqrt{\frac{2}{\pi}} \frac{N^{3/2}}{T^{1/2}}$$

while $E_G$ can be determined from the virial identity [1] $2E_G = E_C + 3(N-1)T$. We note, that the expressions are consistent with the naive scaling behavior of eq. (11), in that both $r_{ee}/N$ and $E_C/N$ are functions of $T/N$ only for large $N$. The relevant small parameter of the high-$T$ expansion is obviously $N/T$. For the blocked system of $K$ effective monomers, the corresponding results are identical to the order shown. Thus, in the unscreened case the high $T$ performance is under control.

In the presence of screening, the situation is somewhat different. The correct and blocked high-$T$ expansions for $E_C$ differ, and it is actually the contribution from the correction term that dominates [5]. Hence, in the screened case the blocking method is reliable only at low $T$.

| N | Q=8 | Q=4 | Q=2 | Q=4/3 |
|---|-----|-----|-----|-------|
| 40 | 1.10 | 1.04 | 1.01 | 1.00 |
| 80 | 1.06 | 1.03 | 1.01 | 1.00 |
| 160 | 1.05 | 1.02 | 1.00 | 1.00 |

Table 1: $r_{ee}(K)/r_{ee}(N)$ as a function of $N$ and $Q = N/K$ at $T=T_r$ for unscreened chains.

We have made extensive numerical evaluations of the blocking approach. All MC runs, both standard and blocked, were performed using the pivot algorithm [2]. Most results are based on $10^4$ thermalization sweeps and $10^5$ measured configurations. Some results were taken from ref. [1]. Initially exploratory comparisons were made in order to find $Q$, $T$ and $\kappa$ ranges where the blocked approach is trustworthy. It turns out that realistic values for $T$ ($T_r$) for unscreened chains are within the domain of application even with very high $Q$-values. As expected, in the presence of screening the blocking approach breaks down when the screening length becomes smaller than the blocked resolution. The results are nevertheless encouraging since even with strong screening ($\kappa$=1.992), $Q \leq 10$ is feasible.

When investigating the power of the method, the relevant parameter is $Q = N/K$. However, in addition one expects finite-$N$ corrections due to end-point effects. This is illustrated in table 1, where the ratios between $r_{ee}$ computed with blocked and standard method, $r_{ee}(K)$ and $r_{ee}(N)$ respectively, are shown for $T=T_r$ and $\kappa$=0.0. As can be seen from table 1 finite $N$ effects diminish as $N$ increases and should not pose problems for large system sizes, $1000 - 10000$ monomers, at which the method is aimed.

We next compare the power of our blocking approach on $r_{ee}$ and $E_C$ for $T = T_r$ and $\kappa$=0.0 and 0.630 respectively. The results for $r_{ee}$ can be found in figs. 1 and 2. The corresponding energies, which very well approximate the "true" MC values, are found in table 2.

| | N | 80 | 160 | 320 | 512 | 1024 | 2048 |
|---|---|----|-----|-----|-----|------|------|
| $\kappa$=0.0 | $E_C$ (K) | 2.46 | 2.76 | 3.04 | 3.22 | 3.48 | 3.72 |
| | (N) | 2.46 | 2.76 | 3.04 | 3.22 | 3.48 | 3.73 |
| $\kappa$=0.63 | $E_C$ (K) | 0.47 | 0.50 | 0.54 | 0.56 | 0.58 | |
| | (N) | 0.47 | 0.48 | 0.48 | 0.48 | 0.48 | |

Table 2: Average internal Coulombic energies per monomer for unscreened and screened Coulomb potentials. N and K stands for full MC and blocked with $K = 50$ respectively.

As can be seen from fig. 1 and table 2 the $r_{ee}$ and $E_C$ are amazingly well described by the blocked approach for the unscreened case. The value of $r_{ee}$ is within 10% from the true value even for $Q \approx 100$. For even larger $N$ the situation should be even better with diminishing finite $N$ effects. Indeed, using the blocked method we have estimated $r_{ee}$ and $E_C$ for a $N$=10000 unscreened chain with $K$=250 and find $r_{ee}/N = 2.04$ and $E_C/N = 4.25$ The relative errors should be less than those quoted for $K = 50$ and $N = 2048$ (3.6% for $r_{ee}$ and 0.3% for $E_C$). The computational speedup here is substantial, growing as $Q^3$.

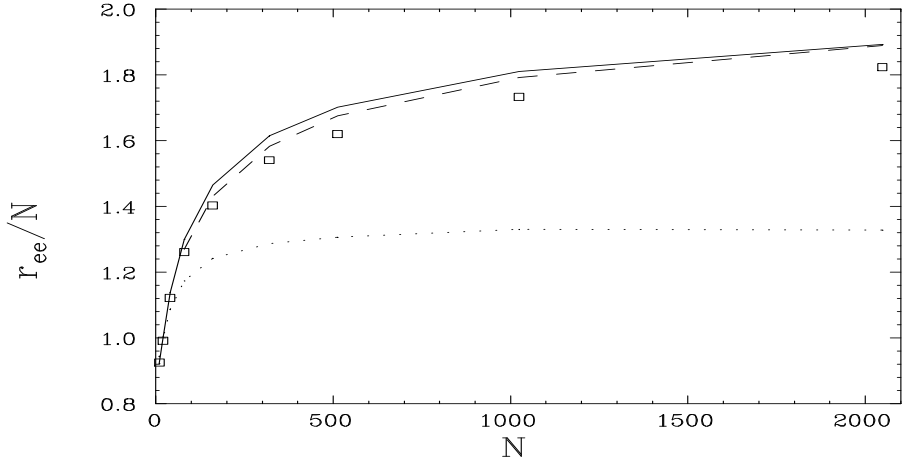

Figure 1: $r_{ee}/N$ as a function of $N$ for an unscreened Coulomb chain. Full and dashed lines denote $K=20$ and $K=50$ blocked MC runs including logarithmic correction respectively. Also shown as dotted lines are the naive blocked results for $K=20$. The squares denote results from original MC runs with $N$ degrees of freedom.

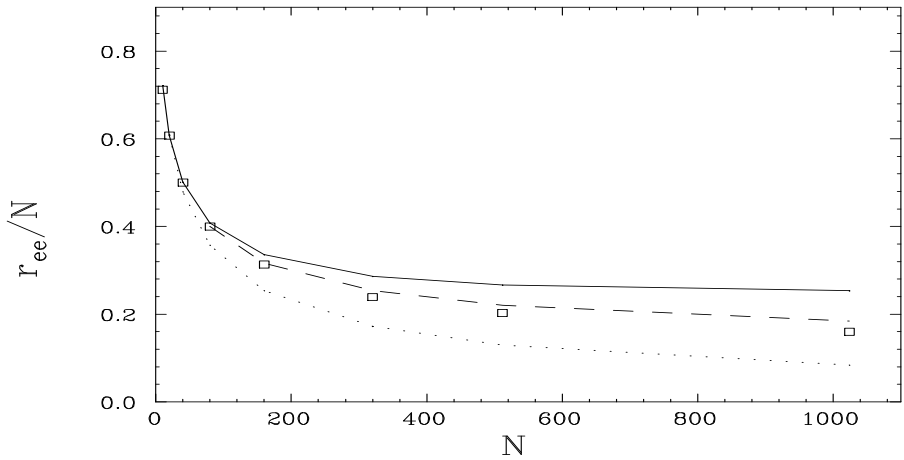

Figure 2: $r_{ee}/N$ as a function of $N$ for a $\kappa=0.630$ screened Coulomb chain. Same notation as in fig. 1.

For the screened case (see fig. 2) the approximation is less accurate but still good enough to be of practical relevance. In table 3 detailed comparisons are made for $Q=4$ and 8 for different screening strengths. As can be seen there, the blocking gives a reasonable approximation even for $Q=8$, giving a speedup factor of 512.

| $\kappa$ | $r_{ee}(N)$ | Q | $r_{ee}(K)$ | % diff | Q | $r_{ee}(K)$ | % diff |
|---|---|---|---|---|---|---|---|
| 0.1992 | 185 | 4 | 186 | 0.6 | 8 | 187 | 1.1 |
| 0.630 | 104 | 4 | 105 | 1.0 | 8 | 110 | 5.8 |
| 1.992 | 61 | 4 | 62 | 1.6 | 8 | 68 | 11.4 |

Table 3: Comparison of $r_{ee}(K,N)$ and $r_{ee}(N,N)$ for $N=512$ screened chains with $Q=4$ and 8.

The approximation is of course not relevant for quantities more local than the blocking allows for.

In summary, we have developed an efficient Monte Carlo blocking scheme that allows for estimating certain thermodynamical quantities of polyelectrolytes with very substantial computational speedup factors for temperatures around $T_r$.. The method is excellent for unscreened chains, but even with screening non-neglible savings result. Results for chain lengths hitherto never probed are reported.

**Acknowledgment:**

We thank Bo Jönsson for valuable comments on the manuscript.

# References

[1] B. Jönsson, C. Peterson and B. Söderberg, *J. Phys. Chem.* **99**, 1251 (1995).

[2] M. Lal, *Mol. Phys.* **17**, 57 (1969);
    N. Madras and A.D. Sokal, *J. Stat. Phys.* **50**, 109 (1988). **50**, 109 (1988).

[3] A. Irbäck, *J. Chem. Phys.* **101**, 1661 (1994).

[4] B. Jönsson, C. Peterson and B. Söderberg, *Phys. Rev. Lett.* **71**, 376 (1994).

[5] C. Peterson, O. Sommelius and B. Söderberg, "Scaling and Scale Breaking in Polyelectrolytes", *LU TP 95-25* (to be submitted to *J. Chem. Phys*).

[6] P.G. deGennes, *Scaling Concepts in Polymer Physics*, Cornell University Press, Ithaca 1979.